\documentclass[graybox]{svmult}

\usepackage{type1cm}        
\usepackage{makeidx}         
\usepackage{graphicx}        
\usepackage{multicol}        
\usepackage[bottom]{footmisc}

\usepackage{newtxtext}       %
\usepackage[varvw]{newtxmath}       

\makeindex             

\begin{document}

\title*{General Relativistic Hydrodynamic Simulations around Accreting Black Holes}
\author{Sudip K Garain\orcidID{0000-0001-9220-0744} and\\ Pranayjit Dey}
\institute{Sudip K Garain \at Department of Physical Sciences and Center of Excellence in Space Sciences India, Indian Institute of Science Education and Research Kolkata, Mohanpur, Nadia 741246, India , \email{sgarain@iiserkol.ac.in}
\and Pranayjit Dey \at Department of Physical Sciences, Indian Institute of Science Education and Research Kolkata, Mohanpur, Nadia 741246, India}
%
%
\maketitle

\abstract*{Strong gravity in the immediate vicinity of 
compact objects (e.g., black holes, neutron stars) 
necessitates inclusion of general relativistic effects. 
Traditionally, pseudo-Newtonian potential representations 
of gravity were favored to simulate the fluid motion in 
this region since that reduced the calculation complexity. 
However, with the advent of easily implementable, reliable 
numerical algorithms and computer hardware, more and more 
research groups are shifting towards the numerical solutions 
of general relativistic fluid dynamics equations. 
In this work, we report our progress on the development of 
such simulation tool and present results of sub-Keplerian 
accretion flow onto black holes.}

\abstract{Strong gravity in the immediate vicinity of 
compact objects (e.g., black holes, neutron stars) 
necessitates inclusion of general relativistic effects. 
Traditionally, pseudo-Newtonian potential representations 
of gravity were favored to simulate the fluid motion in 
this region since that reduced the calculation complexity. 
However, with the advent of easily implementable, reliable 
numerical algorithms and computer hardware, more and more 
research groups are shifting towards the numerical solutions 
of general relativistic fluid dynamics equations. 
In this work, we report our progress on the development of 
such simulation tool and present results of sub-Keplerian 
accretion flow onto black holes.}

\section{Introduction}
\label{sec:1}
Fluid configuration around the black holes determines the 
spectro-temporal and polarimetric
signatures for stellar mass black hole X-ray binaries (BHXRBs) as well 
as the active galactic nuclei (AGNs) \cite{bm2016, netzer2015}. 
Strong gravity around the
black holes mostly dictates the fluid behaviour. Therefore, solution
of general relativistic fluid dynamics equations are frequently used
to infer the spectro-temporal as well as polarimetric properties.

Majority of the general relativistic simulations of accretion disk
start from an initially equilibrium torus threaded with a seed
magnetic field (\cite{porth2019} and references therein).
However a more realistic simulation set up may be constructed 
by letting the matter ideally come from far out and enter the 
simulation domain situated at a finite distance. 
Model based analytical solution for velocity components, 
pressure, density etc. can be supplied as the inflow boundary condition 
at the outer boundary of the simulation domain and 
the time-dependent fluid dynamical equations, with the 
help of this boundary
condition, should determine the dynamical solution inside the domain.
Accretion simulations of collapsing plasma with initially
Bondi type spherically symmetric matter distribution, 
but with latitude-dependent low angular momentum (sub-Keplerian), 
has been done earlier using non-GR codes
(\cite{proga2003,janiuk2008,li2013}) and recently using GRMHD simulations
(\cite{ressler2021,lalakos2022,kaaz2023,cho2023}).
We are, rather, interested in accretion simulations of axisymmetric,
thick disk type configuration.
Such set-up has been used to simulate sub-Keplerian and Keplerian 
matter accretion onto black holes 
\cite{Chakrabarti1993a,Molteni1996b,rcm1997,Giri2013a,Giri2015a,garain2023}. However, all these simulations are done using pseudo-Newtonian
potential proposed by \cite{Paczynsky1980a}.

Our aim is to extend such simulations using general relativistic
fluid dynamics solvers. A few general relativistic hydrodynamics (GRHD)
simulations of transonic, sub-Keplerian accretion disk with this
type of set-up
have been conducted earlier \cite{kgbc2017,kgcb2019}.
However, further extensions are not reported. We are in process
of developing a three dimensional GRHD code, designed specifically
to implement the above set-up, and in this paper, report the initial
results using that code.

Our paper is organized as follows: In Section~\ref{sec:1.5}, 
we provide a very brief overview of the general relativistic
analytical solution of sub-Keplerian accretion flow. 
In Section~\ref{sec:2},
we introduce the GRHD equations and our numerical solution
methodology. In Section~\ref{sec:3}, we present the results.
Finally, in Section~\ref{sec:4}, we provide concluding remarks.

In our following calculations, we use $r_g=GM_{bh}/c^2$ as unit
of distance, $r_g/c$ as unit of time and $r_gc$ as unit of 
specific (i.e., per unit mass) angular momentum. 
Specific energy is measured in the unit of $c^2$.
Here, $G$ is the gravitational constant, $M_{bh}$ is the
mass of the black hole and $c$ is the speed of light in vacuum.

\section{Sub-Keplerian flow : Analytical solutions}
\label{sec:1.5}

The general relativistic analytical solutions for 
transonic, sub-Keplerian accretion
flow onto compact objects are discussed in 
\cite{Chakrabarti1990b,Chakrabarti1996c,Chakrabarti1996b}
with sufficient 
details. A non-dissipative, transonic accretion flow 
is usually characterized 
by the conserved parameters specific energy 
($\epsilon=u_t/(1-na_s^2)$) and specific angular momentum 
($l=-u_\phi/u_t$). Here,
$u_t$ and $u_\phi$ are the $t$ and $\phi$ components of the four-velocity
$u_\mu$, $n$ is the polytropic index and $a_s$ is the local sound speed.

The analytical solution aims to find the radial variation of
the fluid variables under the stationary and axisymmetric conditions.
The calculation starts by considering the equations
of the conserved specific energy $\epsilon=u_t/(1-na_s^2)$ 
and the mass accretion rate $\dot{m}=Au^r\rho$. Here,
$A$ is a geometric quantity representing the surface area through
which mass flux is considered, $u^r$ is the $r$ component of
the four-velocity $u^\nu$ and $\rho$ is the rest mass density.
Further, using the adiabatic equation of state $P=K\rho^\Gamma$,
with $P$ being pressure and $\Gamma$ being the adiabatic index, 
and the relation between enthalpy $h$
and $a_s$ as $h=1/(1-na_s^2)$, $\epsilon$ and $\dot{m}$
are expressed in terms of $a_s$ and 
the radial velocity $\mathcal{V}$ in the
rotating frame. Using the fact that $\epsilon$ and $\dot{m}$ are conserved,
by differentiating these equations w.r.t. $r$ and eliminating
$da_s/dr$, once can obtain an equation for $d\mathcal{V}/dr$
\cite{kgbc2017,kgcb2019}. Numerical integration of this
equation will result in $\mathcal{V}(r)$ and subsequently all other fluid
variables as a function of $r$.


\section{General Relativistic Hydrodynamics: Basic equations and solution procedure}
\label{sec:2}
For numerical simulations, we solve the following conservation equations:
\begin{eqnarray}
\nabla_\mu \left(\rho u^\mu\right) = 0 \\
\nabla_\mu T^{\mu\nu} = 0
\label{eq:01}
\end{eqnarray}
Here, $\nabla_\mu$ represents the covariant derivative,
$u^\mu$ is the four-velocity and 
$T^{\mu\nu}$ is the stress-energy tensor. 
$T^{\mu\nu}=\rho h u^\mu u^\nu + P g^{\mu\nu}$ for ideal fluid
with $h$ as the specific
enthalpy given by $h=1+\frac{\Gamma}{\Gamma -1}\frac{P}{\rho}$, 
$\Gamma=4/3$ being the adiabatic index.
Following \{3+1\} formalism \cite{Banyuls1997a,font2008}, 
we write the space-time metric 
$g_{\mu\nu}$ in terms of lapse ($\alpha$), shift vector
($\beta^i$) and the spatial metric ($\gamma_{ij}$).
After some algebraic manipulations, 
this set of equations can be written as a set of five
partial differential equations (PDEs):
\begin{eqnarray}
\frac{1}{\sqrt{-g}}\left[\frac{\partial \sqrt{\gamma} D}{\partial t} +
		      \frac{\partial}{\partial x^i}
		      \left(\sqrt{-g}D\left( v^i - \frac{\beta^i}{\alpha}\right)\right)\right]&=&0 \\
\frac{1}{\sqrt{-g}}\left[\frac{\partial \sqrt{\gamma} S_j}{\partial t} +
		      \frac{\partial}{\partial x^i}
		      \left(\sqrt{-g}\left(S_j\left( v^i - \frac{\beta^i}{\alpha}\right)+P\delta^i_j\right)\right)\right]&=&
		      T^{\mu\nu}\left(\frac{\partial g_{\nu j}}{\partial x^\mu}-\Gamma^\lambda_{\nu \mu} g_{\lambda j}\right) \\
\frac{1}{\sqrt{-g}}\left[\frac{\partial \sqrt{\gamma} \tau}{\partial t} +
		      \frac{\partial}{\partial x^i}
		      \left(\sqrt{-g}\left(\tau\left( v^i - \frac{\beta^i}{\alpha}\right)+P v^i\right)\right)\right]&=&
		      \alpha \left(T^{\mu 0}\frac{\partial \mathrm{ln}\alpha}{\partial x^\mu} - T^{\mu\nu}\Gamma^0_{\mu\nu}\right)
\end{eqnarray}

Here, $\sqrt{-g}\equiv det(g_{\mu\nu})$ and 
$\sqrt{\gamma}\equiv det(\gamma_{ij})$,
and these are connected by $\sqrt{-g}=\alpha \sqrt{\gamma}$. We denote
the set of five-component vector $U=\left(D, S_j, \tau\right)$ 
as vector of conserved 
variables and can be expressed in terms of vector of primitive
variables $V=\left(\rho, v^i, P\right)$ as follows:
$$
D=\rho W, \quad S_j = \rho h W^2 v_j, \quad \tau = \rho h W^2 - P - D.
$$
Here, $W$ is the Lorentz factor given by 
$W=1/\sqrt{1-v^iv_i}=\alpha u^t$.
$v^i$ are the components of three-velocity given as
$v^i=\frac{u^i}{\alpha u^t} + \frac{\beta^i}{\alpha}$ and the 
co-variant counter part can be calculated as $v_j=\gamma_{ij}v^i$.

The above set of PDEs is further written in integral form and
subsequently discretised on a given mesh \cite{Banyuls1997a,font2008}. 
For our present calculations,
we use Schwarzschild space-time metric for which 
$\alpha=\sqrt{\left(1-2/r\right)}$, $\beta^i=0$, 
$\sqrt{-g}=r^2\sin\theta$ and 
$\sqrt{\gamma}=r^2\sin\theta\left(1-2/r\right)^{-1/2}$.
The resulting discretised equations
on a spherical mesh constructed using Boyer–Lindquist 
coordinates $(t,r,\theta,\phi)$ \cite{Boyer1967a} 
are solved using finite volume method.
A better choice may be to use horizon penetrating Kerr-Schild
coordinate representation of Schwarzschild metric, which we may
adopt in future works.

For spatial reconstruction, we have used second order accurate
van Leer slope limiter following \cite{mignone2014}. We perform 
reconstruction on vector $\left(\rho, Wv^i, P\right)$, instead of
primitive variable vector $V$, since the reconstruction on $Wv^i$
ensures sub-luminal reconstructed profile of $v^i$ inside a zone
\cite{bk2016}. We have provisions for HLL and LLF Riemann solvers
for calculating the interfacial fluxes. Second-order accurate
strong stability preserving Runge-Kutta(RK) time integration is used
for time advancement. One of the non-trivial step in general
relativistic hydrodynamics is conserved-to-primitive conversion
as it requires a non-linear equation solution employing a root
solver (e.g., Newton-Raphson).
We have implemented two methods following \cite{mb2005} and
\cite{del2002}. For our calculations, we prefer the method of
\cite{del2002}. It may happen that the root solver does not converge for
a few zones zones after the time-update step and 
for such pathetic zones, we use the previous
time-step solution as it is already saved in a RK type time-update.
The timestep $dt$ is calculated following standard
Courant-Friedrichs-Lewy (CFL) condition (\cite{leveque2002,toro2009})
$$
dt = C_\mathrm{CFL}\frac{1}{\frac{\lambda^r}{dr} 
                  + \frac{\lambda^\theta}{r d\theta}
		  + \frac{\lambda^\phi}{r\sin\theta d\phi}},
$$
where, $\lambda^i$ is maximum speed in $i^\mathrm{th}$ direction
and $C_\mathrm{CFL}$ is the CFL number. For all the runs, we use
$C_{\mathrm CFL} = 0.9$. For one- or two-dimensional simulations,
contribution from the corresponding inactive dimension(s) is switched
off.

\section{Results}
\label{sec:3}
In this section, we present results of a few standard test 
problems validating our implementation. Later in this section, we present
results for couple of production runs for sub-Keplerian accretion disk.

\subsection{One-dimensional test problems}
In this sub-section, we present results of a couple of test problems
to demonstrate the achievement of global accuracy and correctness
of our code.


\subsubsection{Stationary torus}
\label{subsec:3.1}

In this benchmarking test problem, we initialize a one-dimensional
(radial direction) computation domain [4:40] using a constant 
specific angular momentum ($l=3.9$),
stationary torus solution \cite{abra1978, Chakrabarti1985, font2002}
and evolve the solution for one-full
rotation period $t=100$ at the density maximum. 
Next, we subtract the
numerical solution from the stationary solution and compute the
errors. For a globally (i.e., spatially and temporally) 
second order accurate code, the error should converge with second
order accuracy. For this test problem, we run simulations with
zones ranging from 64 to 4096. The results are shown in Table~\ref{tab:1}.
Both the L$_1$ and L$_\mathrm{inf}$ accuracy columns show achievement
of second order accuracy in the asymptotic limit.

\begin{table}[!t]
\caption{Second order convergence for static torus problem}
\label{tab:1}       
%
%
\begin{tabular}{p{2cm}p{2.4cm}p{2cm}p{2.4cm}p{2cm}}
\hline\noalign{\smallskip}
Zones & L$_1$ error & Accuracy & L$_\mathrm{inf}$ error & Accuracy  \\
\noalign{\smallskip}\svhline\noalign{\smallskip}
64   & 5.11$\times10^{-5}$ &      & 7.14$\times10^{-3}$ &      \\
128  & 1.58$\times10^{-5}$ & 1.69 & 1.99$\times10^{-3}$ & 1.84 \\
256  & 3.72$\times10^{-6}$ & 2.09 & 1.66$\times10^{-3}$ & 0.26 \\
512  & 5.05$\times10^{-7}$ & 2.88 & 2.79$\times10^{-4}$ & 2.57 \\
1024 & 1.06$\times10^{-7}$ & 2.26 & 1.71$\times10^{-5}$ & 4.03 \\
2048 & 2.80$\times10^{-8}$ & 1.92 & 2.84$\times10^{-6}$ & 2.56 \\
4096 & 7.36$\times10^{-9}$ & 1.93 & 7.46$\times10^{-7}$ & 1.93 \\
\noalign{\smallskip}\hline\noalign{\smallskip}
\end{tabular}
\end{table}
%


\subsubsection{Standing accretion and wind shock solutions}
\label{subsec:3.2}

In these one-dimensional benchmarking test problems, 
we demonstrate our code's
capability to capture discontinuous solutions that are predicted
in \cite{fukue1987, chakraba1989}. For the black hole accretion (wind)
solution having multiple sonic points, it is possible that the
solution branch passing through the outer (inner) 
sonic point is connected to
the solution branch passing through the inner (outer) 
sonic point via a shock
transition. Analytically, one needs to perform the Rankine-Hugoniot
analysis to find the shock location. Using numerical simulation,
we are able to detect the shock at the analytically 
predicted location with a satisfactory level of accuracy.

\begin{figure}[h!]
\sidecaption
\includegraphics[scale=.49]{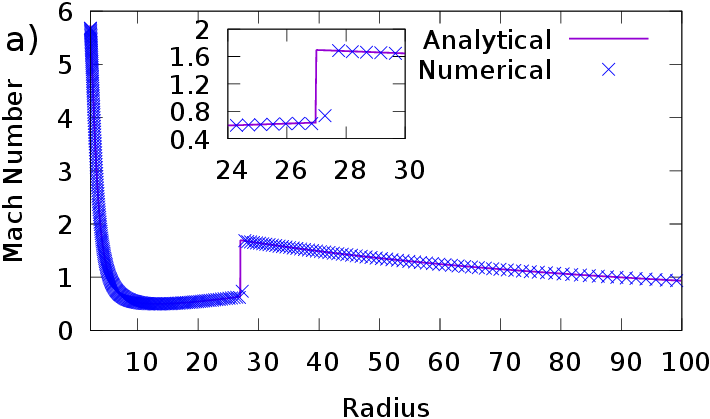}
\includegraphics[scale=.49]{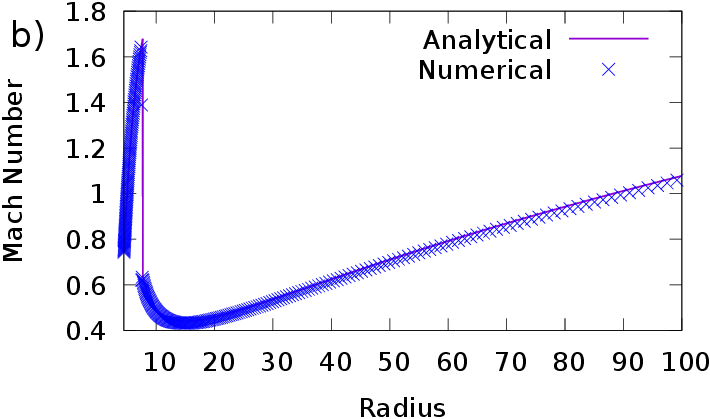}
%
%
\caption{
a) shows the comparison between the analytical (solid line) and the
numerical solutions (crosses) for the shock in accretion solution
whereas b) shows the same for shock in wind solution. Both the plots
show a very good matching between the analytical and numerical
results.
	}
\label{fig:1}       
\end{figure}

Figure~\ref{fig:1}(a) shows an example of the shock in accretion solution.
The solid line shows the analytical solution and the crosses show
the numerical solution. The numerical solution is
shown at a time $t=20000$.
This simulation is run a one-dimensional radial domain [2.2:100]
using 300 ratioed grid (with common ratio = 1.018). We use inflow
type boundary condition at the outer radial boundary $r=100$ and outflow
type boundary condition at the inner radial boundary $r=2.2$.
This solution is specified by the specific
energy ($\epsilon=1.007$) and specific angular momentum ($l=3.4$).
The outer sonic point, shock and the inner sonic points are located 
at 89.6, 27 and 5.4, respectively. The numerical solution clearly
captured all this locations. We notice from Figure~\ref{fig:1}(a)
that flow is highly supersonic when it crosses the inner boundary.
Therefore, acoustic waves do not propagate any feedback upstream 
from the inner boundary.

Figure~\ref{fig:1}(b) shows an example of shock in wind solution.
The line-point styles are same as in Figure~\ref{fig:1}(a). 
Here, the numerical solution is shown at a time $t=30000$.
This simulation is run a one-dimensional radial domain [4.5:100]
using 300 ratioed grid (with common ratio = 1.018). We use inflow
type boundary condition at the inner radial boundary $r=4.5$ and outflow
type boundary condition at the outer radial boundary $r=100$.
This solution is specified by $\epsilon=1.007$ and $l=3.48$. 
The inner sonic point, shock and the outer sonic points are located 
at 5.04, 7.7 and 88.23, respectively. Here again, the numerical 
solution captured all the locations. 


\subsection{Two-dimensional test problems}
In this sub-section, we present results of a couple of test problems
to demonstrate the operability of our code in multi-dimensions.

\subsubsection{Stationary torus}
\label{subsec:3.3}

\begin{figure}[t]
\includegraphics[scale=.55]{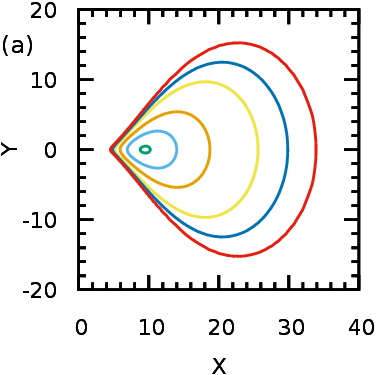}
\includegraphics[scale=.55]{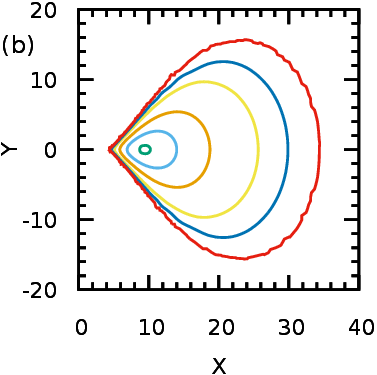}
\includegraphics[scale=.55]{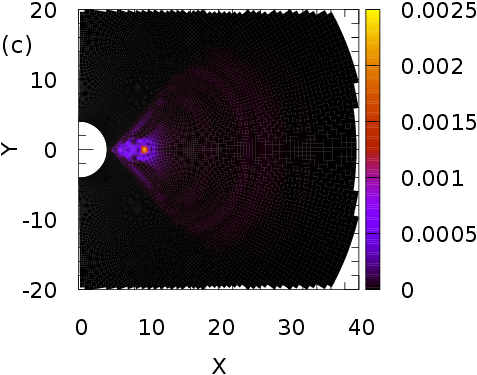}
%
%
\caption{
a) and b) show the contours of constant rest mass densities on log scale
at the initial and the final times, respectively. 
The torus structure is found
to be well maintained except towards the outer layers of the torus. 
c) shows the plot of grid-by-grid difference between the density
values. We see difference is highest around the location of 
density maximum.
	}
\label{fig:2}       
\end{figure}

This is a multi-dimensional extension of the
test problem presented in \ref{subsec:3.1}. We initialize a
constant specific angular momentum
$l=3.9$ stationary torus on the $r-\theta$ domain [4:40]$\times[0:\pi]$
using 128 logarithmically binned radial grids and 180 uniform
angular grids.
The initial condition is evolved till $t=500$ GM/c$^3$
using outflow boundary conditions on radial boundaries and
reflective boundary conditions on polar boundaries. 
Since the torus solution
is a result of hydrostatic equilibrium, it is expected
that the torus structure will be well-maintained except the
numerical dissipation errors. Figure~\ref{fig:2} (a) and (b) show the
contours of constant rest mass densities on log scale
at the initial and the final times, respectively. The inner-most
contour corresponds to log$_{10}$(density)=-1.25 and
successive contours correspond to -1.5, -2, -3, -4 and -7.
The contours
maintain their overall structure unchanged except a few wiggles
on the outer most contour. This is due to the numerical dissipation
of our second-order accurate code. Figure~\ref{fig:2} (c)
shows the grid-by-grid difference of the rest mass density
values and we notice that the maximum difference arises at the
center of the torus, which is the location of density maximum.

\subsubsection{Bondi accretion}
\label{subsec:3.3}

In this test problem, we simulate the spherically symmetric Bondi
accretion \cite{bondi1952} onto a black hole. 
The simulation is performed on 
the $r-\theta$ domain [2.2:100]$\times[0:\pi]$ using 150 
logarithmically binned radial grids and 180 uniform
angular grids. The simulation domain is initially filled
with a background matter having density $10^{-8}$ times
lower than the incoming matter density.
We used inflow boundary condition at the outer radial
boundary and outflow boundary condition at the inner radial
boundary. Thus, at all the ghost zones of the outer radial boundary,
we maintain the vector of primitive variables 
$V=(1.0, -0.05233, 0, 0, 0.005967)$ corresponding
to $\epsilon=1.015$ during the entire simulation.

\begin{figure}[t]
\includegraphics[scale=.7]{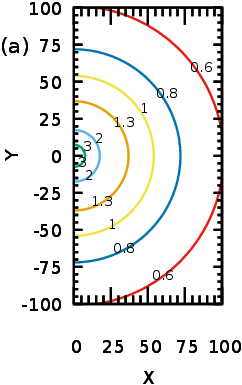}
\includegraphics[scale=.7]{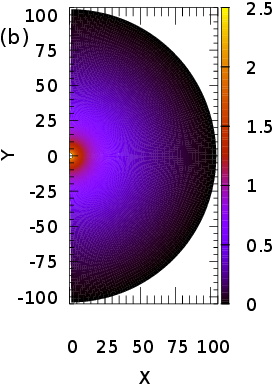}
\includegraphics[scale=.6]{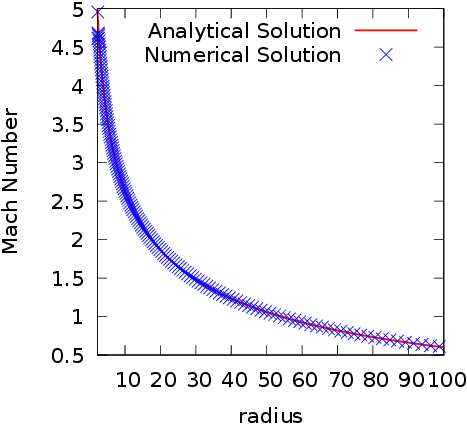}
%
%
\caption{
a) shows the contours of constant Mach numbers whereas b) shows
the rest mass densities on log scale at the final time.
Both these figures demonstrate the spherical symmetry of the solution.
c) shows the comparison of radial Mach number variation between the
analytical (solid line) and the numerical (crosses) results.
        }
\label{fig:3}       
\end{figure}

The simulation is run till the time of $t=10000$. By this time,
a steady state solution is achieved. Figure~\ref{fig:3}(a) shows
the contours of constant Mach numbers whereas (b) shows
the rest mass density distribution on log scale at the final time.
Both these figures demonstrate the spherical symmetry
of the final solution. Figure~\ref{fig:3}(c) shows 
the comparison of radial Mach number variation between the
analytical (solid line) and the numerical (crosses) results.
We find slight mismatch very close to the inner boundary. Here,
the gradients of fluid variables are very steep and we believe,
to capture the correct solution, we need to use either high
order accurate reconstruction schemes or finer
resolution if we continue to use second order accurate schemes.

\subsection{Sub-Keplerian accretion disk}

In this sub-section, we present results for couple of multi-dimensional
simulations of sub-Keplerian accretion disk. We run two cases
with two different $l$ values : for run R1, $l=3.5$ and run R2,
$l=3.56$. $\epsilon = 1.002$ for both the runs. Analytically,
these parameters allow shock formation in the sub-Keplerian
accretion flow. Because of
the higher $l$ value for R2, the average shock location is 
expected to be at higher radial distance.  

For both the cases, the simulations are performed on a 
$r-\theta$ domain [2.1:200]$\times[0:\pi]$ using 300
logarithmically binned radial grids and 180 uniform
angular grids. The simulation domain is initially filled
with a static matter having density and pressure as corresponding
floor values (~ $10^{-8}$ and $10^{-11}$ respectively).
Matter enters the simulation domain at a constant rate through the outer radial boundary.
Velocity components and sound speed of the incoming matter
are calculated following the analytical solution provided 
in Section~\ref{sec:1.5} for a given ($\epsilon$, $l$) pair.
Since we don't have radiative cooling
or viscous dissipations, incoming matter density $\rho_{\mathrm out}$
is normalized to 1.0. The sound speed and $\rho_{\mathrm out}$ together
allow one to calculate the pressure of the incoming matter
for a given $\Gamma$ which is 4/3 for these runs.
Thus, the vector of primitive variables $V$ for the incoming
matter is calculated using ($\epsilon$, $l$) pair and this
$V$ is maintained at the ghost zones of the outer
radial boundary throughout the simulation to mimic constant matter
supply to the black hole.
This inflow boundary condition is maintained for zones having
$80^\circ \leq \theta \leq 110^\circ$.
Other than this, we use outflow boundary condition at all 
other outer radial grids to allow outflow from the accretion disk.
We also use outflow boundary condition at the inner radial
boundary to mimic free-flow of matter towards the horizon. 
Simulations are run till a stopping time of $t=18000$.

\begin{figure}[t]
\includegraphics[scale=.67]{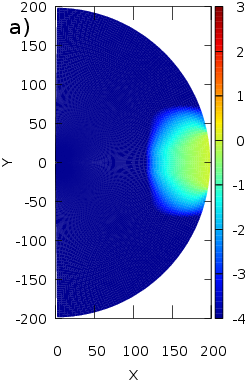}
\includegraphics[scale=.67]{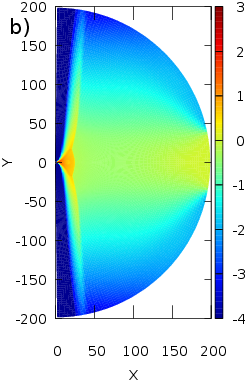}
\includegraphics[scale=.67]{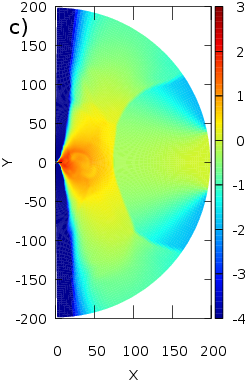}
\includegraphics[scale=.67]{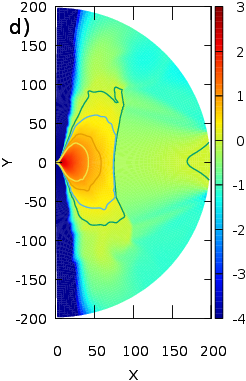}
%
%
\caption{
shows the time evolution of the sub-Keplerian 
accretion disk structure. Colors
show the rest mass density distribution. Snapshots are shown
at times a) 400, b) 1600, c) 6000
and d) 18000. Density isocontours corresponding to density
values 10, 5, 2 and 1 as we move from inner-most contour
to the outer one, are over-plotted in d). The isocontours
clearly show the formation of density torus in the post-shock region.
        }
\label{fig:4}       
\end{figure}

\begin{figure}[t]
\sidecaption
\includegraphics[scale=.5]{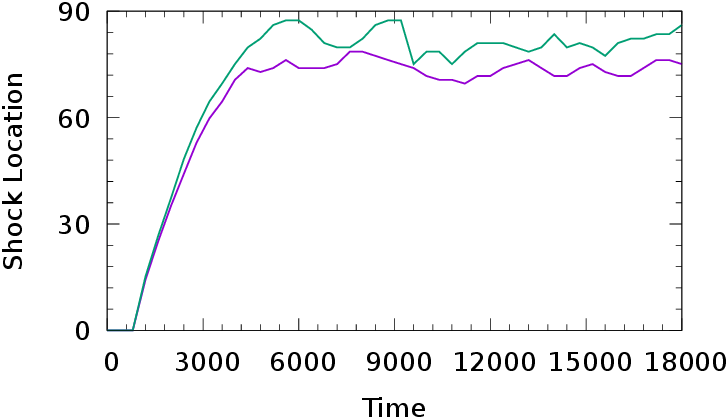}
%
%
\caption{ 
shows the time variation of the shock location on the equatorial
plane for the two different runs (green - R2, purple - R1). 
This plot shows that the post-shock
torus is dynamic rather than being static.
        }
\label{fig:5}       
\end{figure}

Figure~\ref{fig:4} shows the time-evolution for run R1. Colors
show the density distribution at times a) 400, b) 1600, c) 6000
and d) 18000. Time evolution for run R2 follows similar
pattern. The simulations achieve
a nearly steady state around time $t=6000$ (i.e., state
corresponding to Fig.~\ref{fig:4}(c). 
Rest mass density isocontours corresponding to density
values 10, 5, 2 and 1 as we move from inner-most contour
to the outer one, are over-plotted in Fig.~\ref{fig:4}(d). The isocontours
clearly show the formation of density torus in the post-shock region.
The density torus resembles the thick torus which are constructed
using hydrostatic equilibrium assumption (e.g., Figure~\ref{fig:2}).
However, our simulated torus has advection included and is highly
dynamic. The former is a result of balance between the 
inward gravitational force
and outward combined effect of centrifugal and pressure gradient forces.
The flow is purely azimuthal with radial component of four velocity
set to zero. However, in our case, the flow has both azimuthal and radial
non-zero velocity components. Because of this radial component,
energy-momentum is advected towards the black hole.
The outer boundary of the torus coincides with the
location of shock in the accreting sub-Keplerian matter.
Figure~\ref{fig:5} shows the time variation of the shock 
location on the equatorial plane for the two different runs
(green - R2, purple - R1). This plot shows that the post-shock
torus is dynamic rather than being static. Also, the torus
size is larger for higher $l$.
Such post-shock dynamical tori are used to explain the
observed spectral and temporal properties of accreting black holes
\cite{cui1997,cm2000,radhika2016,shang2019}.

\section{Summary and Conclusions}
\label{sec:4}

In this paper, we present our progress of developing a general
relativistic fluid dynamics solver. Our aim is to use the
said solver for simulating an accretion disk configuration
that mimics mass inflow from far out rather than starting 
from an equilibrium torus. We have demonstrated
that our presently developed GRHD code works for Schwarzschild 
spacetime, is globally second 
order accurate and performs robustly in multi-dimensions.
Finally, using this code, we simulate geometrically
thick sub-Keplerian accretion disks.
At the time of writing this paper, we have extended
the code's operability in three-dimensions and the performance
is being tested. We'll report the results in future publications.
%
%
%
\begin{acknowledgement}
We acknowledge the usage of Kepler cluster of DPS, IISER Kolkata
and Pegasus cluster of IUCAA, Pune for running a few simulations.
SKG also acknowledges the support of start-up research grant provided by
IISER Kolkata. The authors thank the anonymous referee for constructive 
suggestions for improving the manuscript.
\end{acknowledgement}
\ethics{Competing Interests}{
The authors have no conflicts of interest to declare that are relevant to the content of this chapter.}


\bibliographystyle{spphys}
\bibliography{references}
\end{document}